\documentstyle[12pt]{article}
\begin{document}
\tolerance=5000
\def\be{\begin{equation}}
\def\ee{\end{equation}}
\def\bea{\begin{eqnarray}}
\def\eea{\end{eqnarray}}
\def\nn{\nonumber \\}
\def\cF{{\cal F}}
\def\det{{\rm det\,}}
\def\Tr{{\rm Tr\,}}
\def\e{{\rm e}}
\def\etal{{\it et al.}}

\  \hfill 
\begin{minipage}{3.5cm}
NDA-FP-33 \\
May 1997 \\
\end{minipage}

\ 

\vfill

\begin{center}

{\Large\bf Trace Anomaly and Non-Local Effective 
Action for 2D Conformally Invariant Scalar Interacting with 
Dilaton}

\vfill

{\large\sc Shin'ichi NOJIRI}\footnote{
e-mail : nojiri@cc.nda.ac.jp}
and
{\large\sc Sergei D. ODINTSOV$^{\spadesuit}$}\footnote{
e-mail : odintsov@quantum.univalle.edu.co}

\vfill

{\large\sl Department of Mathematics and Physics \\
National Defence Academy \\
Hashirimizu Yokosuka 239, JAPAN}

{\large\sl $\spadesuit$ Dep.de Fisica \\
Universidad del Valle \\
AA25360, Cali, COLUMBIA \\
and \\
Tomsk Pedagogical Univer. \\
634041 Tomsk, RUSSIA}

\vfill

{\bf ABSTRACT}

\end{center}

Using the results of the calculation of the one-loop 
effective action (E. Elizalde \etal , {\sl Phys.Rev.}
{\bf D49} (1994) 2852), we find the trace anomaly 
for most general conformally invariant 2D dilaton 
coupled scalar-dilaton system (the contribution 
of dilaton itself is included).
The non-local effective action induced by conformal 
anomaly for such system is found.
That opens new possibilities in generalizing of 
CGHS-like model for the study of back reaction of 
matter to 2D black holes.

\newpage

\noindent
1. In the study of back reaction of quantum conformal 
matter to the evaporation of 2D black holes, 
the trace anomaly plays an important role.
For example in the seminal works \cite{CGHS} and 
\cite{RST} working in the large-$N$ approximation 
(where $N$ is the number of scalars minimally coupled 
to dilaton), such trace anomaly has been used for the calculation of anomaly-induced effective action.
This action represents back-reaction of matter to 
dilaton gravity theory and it may significally 
change the properties of 2D black holes and 
Hawking radiation \cite{CGHS,RST}.
Moreover, such trace anomaly is directly responsible 
for Hawking radiation.

In more general dilaton gravity models (for such 
models, see \cite{SOB}) where scalar non-minimally 
couples with dilaton the trace anomaly is more 
complicated.
That is this trace anomaly which should be used 
in the generalizing of CGHS model where 
back-reaction is given by dilaton-coupled scalar.
Few days ago, the calculation of trace anomaly 
for dilaton coupled scalar has been presented in 
ref.\cite{HB} (for a very specific choice of 
dilaton coupling). 
The purpose of the present note is to find the 
trace anomaly for most general conformally 
invariant dilaton-scalar system.
Actually, such result is the by-product of the 
calculation of the one-loop effective action 
for correspondent system which was done sometime 
ago in ref.\cite{ENO}.
We also find the anomaly-induced effective action 
for corespondent system.

\noindent
2. Let us calculate trace anomaly for the system of 
scalar fields interacting with dilaton.
In this calculation, we will take into account also 
the contribution of dilaton to trace anomaly.

The most general action to start with is given by 
\be
\label{I}
S=\int d^2x \sqrt g \left\{ -{1 \over 2}Z(\phi) 
g^{\mu\nu}\partial_\mu\phi\partial_\nu\phi 
+ {1 \over 2}
f(\phi) g^{\mu\nu}\partial_\mu\chi_i
\partial_\nu\chi_i
\right\}
\ee
where $\phi$ is dilaton, $Z$, $f$ are the arbitrary 
dilaton functions, $\chi$ is scalar field, 
$i=1,\cdots,N$.
Here dilaton and scalars are quantum fields and 
gravitational field is considered to be external 
field.

It is clear that such theory is conformally invariant 
in two dimensions.
Hence, on classical level the trace of 
energy-momentum tensor is zero, while on quantum 
level, it is not zero (conformal or trace anomaly).

We work on the background with non-zero background 
dilaton and non-zero background scalar.
The calculation of the one-loop effective action for 
such system has been done in 
all details, in ref.\cite{ENO} (see Eq.(41)).
The result looks as follows:
\bea
\label{eno-result}
\Gamma_{{\rm div}}
&=&- {1 \over 4\pi(n-2)}\int d^2x \sqrt g 
\Bigl\{ -{N+1 \over 6}R + 
\left({{f'}^2 \over 2fZ}-{f'' \over 2Z}\right) 
(\nabla^\lambda \chi_i)(\nabla_\lambda \chi_i) \nn
&& + \left({N f'' \over 2f}-{N{f'}^2 \over 4f^2} 
- {{Z'}^2 \over 4 Z^2} \right)
(\nabla^\lambda \phi)(\nabla_\lambda \phi) 
+\left({Nf' \over 2f}-{Z' \over 2Z}\right)
\Delta \phi \Bigr\}
\eea
Here $\phi$, $\chi_i$ denote the dilaton and scalar 
background correspondently.

It is widely well-known that one can easily get the 
trace anomaly of conformally invariant system just 
by calculating the corresponding divergent 
effective action multiplied to $(n-2)$ in 2 
dimensions (in other words, as coefficient of the 
pole). 
Hence, the trace anomaly of the system under 
discussion is given by
\bea
\label{trace}
T&=&{1 \over 24\pi}\Bigl\{ (N+1)R
-3\left({{f'}^2 \over fZ}-{f'' \over Z}\right) 
(\nabla^\lambda \chi_i)(\nabla_\lambda \chi_i) \nn
&& - 3\left({N f'' \over f}-{N{f'}^2 \over 2f^2} 
- {{Z'}^2 \over 2 Z^2} \right)
(\nabla^\lambda \phi)(\nabla_\lambda \phi) 
- 3\left({Nf' \over f}-{Z' \over Z}\right)
\Delta \phi \Bigr\}
\eea
For purely scalar system coupled with dilaton 
the trace anomaly looks like:
\bea
\label{trace2}
T&=&{1 \over 24\pi}\Bigl\{ NR
 - 3N\left({ f'' \over f}-{{f'}^2 \over 2f^2} \right)
(\nabla^\lambda \phi)(\nabla_\lambda \phi) \nn
&& - 3{Nf' \over f}\Delta \phi \Bigr\}
\eea
Hence, we got the trace anomaly for scalar coupled 
with dilaton and for dilaton itself as by-product 
of the calculation of the effective action in 
ref.\cite{ENO}.

Let us consider very special case $N=1$ (single 
scalar) and $f(\phi)=\e^{-2\phi}$. 
Then
\be
\label{trace3}
T={1 \over 24\pi}\Bigl\{ R
- 6(\nabla^\lambda \phi)(\nabla_\lambda \phi) 
+6\Delta \phi \Bigr\}
\ee
The calculation of conformal anomaly has been done 
recently in the last case, using zeta-regularization 
method (see \cite{E} for a review) in the ref.\cite{HB}.
Comparing our expression (\ref{trace3}) with 
corresponding expression of ref.\cite{HB}, we 
see the difference in coefficient of the last 
term (total divergence) in Eq.(\ref{trace3}), 
which is denoted by $q_3$ in ref.\cite{HB}.
The disagreement is presumably caused by the fact 
that in getting of expression (\ref{eno-result}), 
 one could drop some total divergence term as we are only 
interesting in the dilaton dependence of trace 
anomaly.
In fact, the last term in eq.(3.3) in ref.\cite{HB} 
is, as written, is not unique and the term can be
\be
\label{gene33}
q_3\left\{a\phi R + (1-a) \Box \phi {1 \over \Box}R
\right\}
\ee
or more general form. 
Here $a$ is an arbitrary or undetermined constant.
The authors of ref.\cite{HB} claimed 
that the difference is total derivative 
and can be neglected but it would not be true.
As in section 3.1 in ref.\cite{HB}, if we 
consider the case where $\phi$ is a constant 
$\phi=\phi_0$, we obtain 
\be
\label{cor34}
q_3 a \phi_0 R
\ee
instead of the last term in eq.(3.4) of ref.\cite{HB}.
This tells $q_3$ cannot be determined due to the 
undetermined constant $a$. It indicates that the coefficient 
of total divergence term in the trace can not be derived in unique way.
The coefficients of first two terms in (\ref{trace3}) 
coincide with the calculation of ref.\cite{HB}.

Thus we got the trace anomaly for most general 2D 
dilaton coupled scalar-dilaton system with 
arbitrary dilaton interactions.

\noindent
3. Using the trace anomaly obtained in the previous 
section (see eq.(\ref{trace}),  we may calculate the non-local effective action induced by trace anomaly.

Let us write the general form of trace anomaly as following
\bea
\label{trace4}
T&=&cR + F_1(\phi)
(\nabla^\lambda \chi_i)(\nabla_\lambda \chi_i) + F_2(\phi)(\nabla^\lambda \phi)(\nabla_\lambda \phi) 
\nn && +F_3(\phi)\Delta \phi 
\eea
where the explicit form of $c$, $F_1$, $F_2$, $F_3$ is 
evident from the comparison with (\ref{trace3}):
\bea
\label{comp}
&& c={N+1 \over 24\pi}\ ,\ \ 
F_1(\phi)=-{1 \over 8\pi}
\left({{f'}^2 \over fZ}-{f'' \over Z}\right), \nn 
&& F_2(\phi)=-{1 \over 8\pi}
\left({N f'' \over f}-{N{f'}^2 \over 2f^2} 
- {{Z'}^2 \over 2 Z^2} \right)\ ,  \nn
&& F_3(\phi)=-{1 \over 8\pi}
\left({Nf' \over f}-{Z' \over Z}\right)\ .
\eea
The (non-local) effective action induced by the 
conformal anomaly is defined as 
\be
\label{effaction}
\int d^2x\sqrt g T = 2 \left.{dW \over dt}
\right|_{t=1}
\ee
where $\tilde g^{\mu\nu}=t^{-1}g_{\mu\nu}$. 
In other words, the effective action (\ref{effaction}) 
should derive the anomaly $T$ (see \cite{SO}).

First of all, let us rewrite the trace anomaly 
in the different form.
To do this, first of all, we introduce the function 
$\tilde F_3(\phi)$ such that 
${\partial \tilde F_3(\phi) \over \partial \phi}
=F_3(\phi)$. Then 
$\tilde F_3(\phi)=\int F_3(\phi) d\phi$ and 
\bea
\label{trace5}
T&=&cR + F_1(\phi)
(\nabla^\lambda \chi_i)(\nabla_\lambda \chi_i) \nn
&& + \left(F_2(\phi)- {\partial F_3(\phi) \over 
\partial \phi}\right)
(\nabla^\lambda \phi)(\nabla_\lambda \phi) 
 +\Delta \tilde F_3(\phi) 
\eea
At the next step, we introduce the function 
$\tilde F_2(\phi)$ where ${\partial \tilde 
F_2(\phi) \over \partial \phi}
=\left(F_2(\phi)- {\partial F_3(\phi) \over 
\partial \phi}\right)^{1 \over 2}$.
Then
\bea
\label{trace6}
T&=&cR + 
(\nabla^\lambda \tilde\chi_i)
(\nabla_\lambda \tilde\chi_i) \nn
&& + \nabla^\lambda \tilde F_2(\phi)
\nabla_\lambda \tilde F_2(\phi) 
 +\Delta \tilde F_3(\phi) 
\eea
where $\nabla^\lambda\tilde\chi_1\equiv F_1^{1 \over 2}
(\phi)\nabla^\lambda \chi_i$.

The non-local effective action induced by the trace 
anomaly of such form has been already obtained in 
ref.\cite{HB}.
Applying this result to Eq.(\ref{trace6}) we get
\bea
\label{indaction}
W&=&-{1 \over 2}\int d^2x \sqrt g \Bigl[ 
{c \over 2}R{1 \over \Delta}R + 
(\nabla^\lambda \tilde\chi_i)
(\nabla_\lambda \tilde\chi_i) {1 \over \Delta}R \nn
&& + \nabla^\lambda \tilde F_2(\phi)
\nabla_\lambda \tilde F_2(\phi) {1 \over \Delta}R 
+ \tilde F_3(\phi) R
\eea
One can rewrite this efective action in terms of 
original functions;
\bea
\label{indaction2}
W&=&-{1 \over 2}\int d^2x \sqrt g \Bigl[ 
{c \over 2}R{1 \over \Delta}R + F_1(\phi)
(\nabla^\lambda \chi_i)(\nabla_\lambda \chi_i) 
{1 \over \Delta}R \nn
&& + \left(F_2(\phi)- {\partial F_3(\phi) \over 
\partial \phi}\right)\nabla^\lambda \phi
\nabla_\lambda \phi {1 \over \Delta}R 
+ R\int F_3(\phi) d\phi \Bigr]
\eea
Thus, we got the trace anomaly induced action for 
most general 2D dilaton coupled scalar-dilaton 
system.
This action should be added to the classical action of 
dilaton gravity interactng with above system in order 
to study the properties of black holes and Hawking 
radiation with account of back-reaction.
We hope to return to the study of applications of 
the effective action (\ref{indaction2}) in the 
evaporation of 2D black holes taking into account 
quantum effects in near future.

\end{document}